\documentclass[aps,prl,twocolumn,superscriptaddress]{revtex4-1}

\usepackage{graphicx}
\usepackage{dcolumn}
\usepackage{bm}
\usepackage{amsmath}
\usepackage{amssymb}

\begin{document}


\title{Enhanced Superconductivity and Suppression of Charge-density Wave Order\\ in 2H-TaS$_2$ in the Two-dimensional Limit
}


\author{Yafang Yang}
\affiliation{Department of Physics, Massachusetts Institute of Technology, Cambridge, MA 02139 USA}
\author{Shiang Fang}
\affiliation{Department of Physics, Harvard University, Cambridge, Massachusetts 02138, USA}
\author{Valla Fatemi}
\affiliation{Department of Physics, Massachusetts Institute of Technology, Cambridge, MA 02139 USA}
\author{Jonathan Ruhman}
\affiliation{Department of Physics, Massachusetts Institute of Technology, Cambridge, MA 02139 USA}
\author{Efr{\'e}n Navarro-Moratalla}
\affiliation{Instituto de Ciencia Molecular, Universidad de Valencia, c/Catedr{\'a}tico Jos{\'e} Beltr{\'a}n 2, 46980 Paterna, Spain}
\author{Kenji Watanabe}
\affiliation{National Institute for Materials Science, Namiki 1-1, Tsukuba, Ibaraki 305-0044, Japan}
\author{Takashi Taniguchi}
\affiliation{National Institute for Materials Science, Namiki 1-1, Tsukuba, Ibaraki 305-0044, Japan}
\author{Efthimios Kaxiras}
\affiliation{Department of Physics, Harvard University, Cambridge, Massachusetts 02138, USA}
\affiliation{John A. Paulson School of Engineering and Applied Sciences, Harvard University, Cambridge, Massachusetts 02138, USA}
\author{Pablo Jarillo-Herrero}
\email{pjarillo@mit.edu}
\affiliation{Department of Physics, Massachusetts Institute of Technology, Cambridge, MA 02139 USA}


\date{\today}

\begin{abstract}

As superconductors are thinned down to the 2D limit, their critical temperature $T_c$ typically decreases. 
Here we report the opposite behavior, a substantial enhancement of $T_c$ with decreasing thickness, in 2D crystalline superconductor 2H-TaS$_2$. 
Remarkably, in the monolayer limit, $T_c$ increases to 3.4 K compared to 0.8 K in the bulk. 
Accompanying this trend in superconductivity, we observe suppression of the charge-density wave (CDW) transition with decreasing thickness. 
To explain these trends, we perform electronic structure calculations showing that a reduction of the CDW amplitude results in a substantial increase of the density of states at the Fermi energy, which contributes to the enhancement of $T_c$.
Our results establish ultra-thin 2H-TaS$_2$ as an ideal platform to study the competition between CDW order and superconductivity.
\end{abstract}


\maketitle


Transition metal dichalcogenides (TMDs) 2H-MX$_2$ (where M = Nb, Ta and X = S, Se) have attracted considerable attention as intriguing 2D crystalline superconductors \cite{Saito:2016ht}. 
In these materials, superconductivity (SC) forms in an environment of pre-existing charge-density wave (CDW) order \cite{neto2001charge,kiss2007charge}, making it an ideal platform to study many-body ground states and competing phases in the 2D limit.
In bulk crystals, the reported critical temperature of the CDW transition decreases from 120 K in 2H-TaSe$_2$ down to 30 K in 2H-NbSe$_2$. 
Superconductivity weakens in approximately reverse order, with $T_c$ increasing from around
0.2 K in 2H-TaSe$_2$ to 7.2 K in 2H-NbSe$_2$.
The relationship between CDW and superconductivity in such systems is still under debate \cite{calandra2009effect,ge2012effect}.
It is generally believed that such mutual interaction is competitive, but evidence to the contrary, indicating a cooperative interaction, has also been reported in angular-resolved photoemission spectroscopy studies \cite{kiss2007charge}.

In TMDs, superconductivity and CDW instability can be investigated by adjusting the interlayer interactions through pressure \cite{freitas2016strong,chu1977pressure} or molecule intercalation \cite{thompson1972effects,wilson1975charge}.
Recently, mechanical exfoliation has emerged as a robust method for producing ultra-clean, highly crystalline samples with atomic thickness \cite{geim2007rise}.
This offers a useful way to assess the effect of dimensionality and interlayer interactions on superconductivity and CDW.
A material whose behavior as a function of layer thickness has been recently studied is 2H-NbSe$_2$ \cite{cao2015quality,xi2016ising,Tsen:2015dw}, in which the superconducting state is progressively weaker in thinner samples, with $T_c$ reduced from 7 K in bulk crystals to 3 K in the monolayer.
The thickness dependence of CDW order is still under debate, considering discrepancies between Raman and scanning tunneling microscopy/spectroscopy (STM/STS) studies
\cite{xi2015strongly, Ugeda:2015fu}.

Bulk 2H-TaS$_2$, another member of the 2H-MX$_2$ family, exhibits a CDW transition at 70 K and a SC transition at 0.8 K \cite{nagata1992superconductivity, neto2001charge, Guillamon:2011bz, NavarroMoratalla:2016df}.
Compared to 2H-NbSe$_2$, 2H-TaS$_2$ manifests a stronger signature of CDW transition in transport in the form of a sharp decrease of resistivity \cite{wilson1975charge}, and thus serves as a desirable platform to study the thickness dependence of CDW instability.
STM/STS measurements on monolayer TaS$_2$ epitaxially grown on Au(111) substrates show suppression of the CDW instability \cite{sanders2016crystalline}.
Recently, a study observed an enhanced $T_c$ down to a thickness of 3.5 nm, utilizing TaS$_2$ flakes directly exfoliated on a Si/SiO$_2$ substrate \cite{NavarroMoratalla:2016df}.
Unfortunately, it was found that samples thinner than that become insulating, indicative of its particular susceptibility to degradation in ambient atmosphere. 
Therefore, exfoliation and encapsulation in an inert atmosphere become crucial to attain high quality samples.

Here we report that superconductivity persists in 2H-TaS$_2$ down to the monolayer limit, with a pronounced increase in $T_c$ from 0.8 K in bulk crystals to 3.4 K in the monolayer.
Two transport observations, that in the bulk signal the CDW transition, are found to vanish in ultra-thin samples: (1) a kink in the temperature dependence of the resistivity, and (2) a change of sign in the Hall coefficient versus temperature.
In search of an origin for such trends, we perform electronic structure and phonon spectrum calculations. 
We show that suppression of the CDW order leads to a substantial increase in the density of states at the Fermi level, $N(E_F)$, which ultimately enhances $T_c$.
Our observations also motivates consideration of quantum fluctuations of the CDW order as another mechanism that boosts $T_c$. 
This provides new insights into the impact of reduced dimensions on many-body ground states and their interactions. 

\begin{figure}
\includegraphics[width=3.365 in]{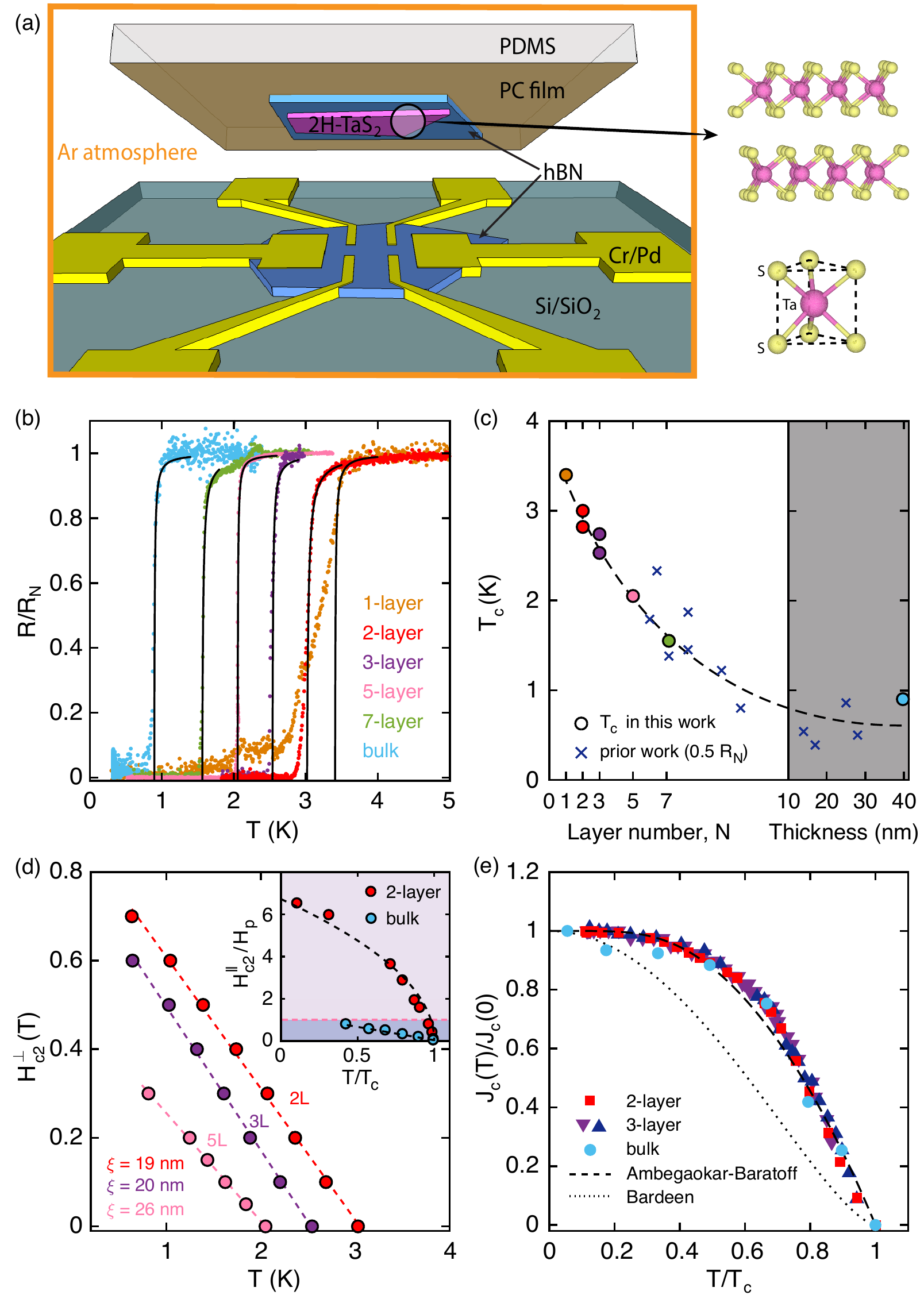}
\caption{
(a) Schematic of device fabrication and crystal structure of 2H-TaS$_2$. (b) Resistance normalized by the normal state ($R/R_N$) as a function of temperature for 1, 2, 3, 5, 7-layer and bulk ($d=40$ nm) samples near the SC transition. The superconducting $T_c$ is 3.4, 3.0, 2.5, 2.05, 1.6 and 0.9 K respectively, determined by fitting the transition curve to the Aslamazov-Larkin formula (black solid lines). (c) $T_c$ reported in this work (circles) and in a prior study (crosses) \cite{NavarroMoratalla:2016df}. The dashed line guides the eye to the general trend. (d) Out-of-plane critical field $H_{c2}$ for 2, 3, 5-layer samples. The dashed lines are linear fits to $H_{c2}^{\perp}= \phi_0/(2 \pi \xi(0)^2)(1-T/T_{c})$, where $\phi_0$, $\xi(0)$ denote the flux quantum and in-plane GL coherence length at zero temperature respectively. Inset: In-plane critical field $H_{c2}^{\parallel}$ normalized by Pauli limit ($H_p \approx 1.86 T_c$) for bilayer and bulk samples. The dashed line for bilayer is a fit to the Tinkham formula for 2D samples $H_{c2}^{\parallel}= \sqrt{12} \phi_0/(2 \pi \xi(0) d) \sqrt{(1-T/T_{c})}$ \cite{tinkham1996introduction}. The purple background indicates the Pauli limit regime. (e) Normalized critical current as a function of $T/T_c$. Dashed and dotted lines denote the models proposed by Bardeen \cite{bardeen1962critical} and Ambegaokar-Baratoff \cite{ambegaokar1963tunneling} respectively.
}
\label{fig1}
\end{figure}

In this work, we exfoliate and fabricate samples with a transfer set-up built inside a glove box filled with Argon gas, and encapsulate the TaS$_2$ flake between two sheets of hexagonal boron nitride \cite{SM}. 
We build our devices utilizing a polymer pick-up technique \cite{wang2015electronic} as illustrated in Fig.~\ref{fig1} (a), taking advantage of the van der Waals adhesion between 2D layers.
With this method, we are able to obtain high quality few-layer TaS$_2$ devices.
As seen in Fig.~\ref{fig1} (b), when temperature is sufficiently low, a clear superconducting transition is observed for 1, 2, 3, 5, 7-layer, and bulk samples.
By fitting the resistance to the Aslamazov-Larkin expression \cite{aslamasov1968influence}, we are able to determine the mean-field superconducting transition temperature $T_c$. 
When sample thickness is decreased from bulk to monolayer, the corresponding $T_c$ monotonically increases from 0.9~K to 3.4~K \footnote{For the monolayer flake, determination of $T_c$ becomes tricky due to electrical shortage to adjacent flakes with different thickness. Detailed analysis and determination of $T_c$ by critical current mapping can be found in Supplemental Material \cite{SM}.}.
The trend observed in our experiment is strictly opposite that of a previous finding on 2H-NbSe$_2$ \cite{xi2016ising}, where $T_c$ decreases monotonically with thickness reduction, despite the fact that the two materials are isostructural and isovalent.
In 2H-NbSe$_2$, the decreased $T_c$ is attributed to a weaker interlayer Cooper pairing given by $\lambda_{inter} \sim$ cos$(\pi/(N+1))$ as layer number $N$ is reduced. 
In our case, however, it is surprising to see that even with a reduced interlayer Cooper pairing, the $T_c$ is dramatically enhanced. 
To verify that the thickness dependence of $T_c$ is independent of extrinsic factors (such as level of disorder, substrate, source of crystal, sample quality), we measure an additional set of bilayer and trilayer samples and plot our results alongside previously reported $T_c$ values \cite{NavarroMoratalla:2016df} in Fig.~\ref{fig1} (c).
Regardless of different sample preparation procedures and substrates, the trend of $T_c$ versus thickness is consistent between the two sets of experimental results, indicative of an intrinsic origin underlying the enhancement of $T_c$.

To further characterize the superconducting properties of thin TaS$_2$, we characterize the superconducting transition under both out-of-plane and in-plane magnetic fields.
Fig.~\ref{fig1} (d) shows the $H_{c2}^{\perp}$ as a function of temperature. 
Close to $T_c$,  the dependence of $H_{c2}^{\perp}$ is fitted well by the phenomenological 2D Ginzburg-Landau (GL) model, which yields $\xi(0)=19,~20,~26$ nm for 2, 3, 5-layer respectively, where $\xi(0)$ denotes the GL coherence length at zero temperature.
The reported value for 3D ranges from 22 nm \cite{kashihara1979upper} to 32.6 nm \cite{abdel2016enhancement}.
For in-plane field, we observe a much larger $H_{c2}^{\parallel} = 32$ T at 300 mK for a bilayer ($T_c = 2.8$ K), which is more than six times the Pauli paramagnetic limit $H_p$, obeying a square root rather than a linear temperature dependence (inset of Fig.~\ref{fig1} (d)).
This dramatic enhancement of in-plane $H_{c2}^{\parallel}$, often referred to as Ising superconductivity, has been observed in other 2D crystalline superconductors \cite{lu2015evidence, saito2016superconductivity,xi2016ising,Tsen:2015dw}. 
The above observations verify that thin TaS$_2$ behaves as a 2D superconductor. 
We also show that the superconducting transition also exhibits the Berezinskii-Kosterlitz-Thouless (BKT) transition as expected in 2D in the Supplemental Material \cite{SM}.
Additionally, the critical current density increases by orders of magnitude as the devices become thinner (bulk $J_c\approx 700$ A/cm$^{2}$, trilayer $J_c\approx 7\times10^5$ A/cm$^{2}$, bilayer $J_c\approx 1.2\times10^6$ A/cm$^{2}$).
The trend of critical current density versus temperature for representative thicknesses is shown in Fig.~\ref{fig1} (e), with a more detailed discussion in Supplemental Material \cite{SM}.

\begin{figure}
\includegraphics[width=3.375 in]{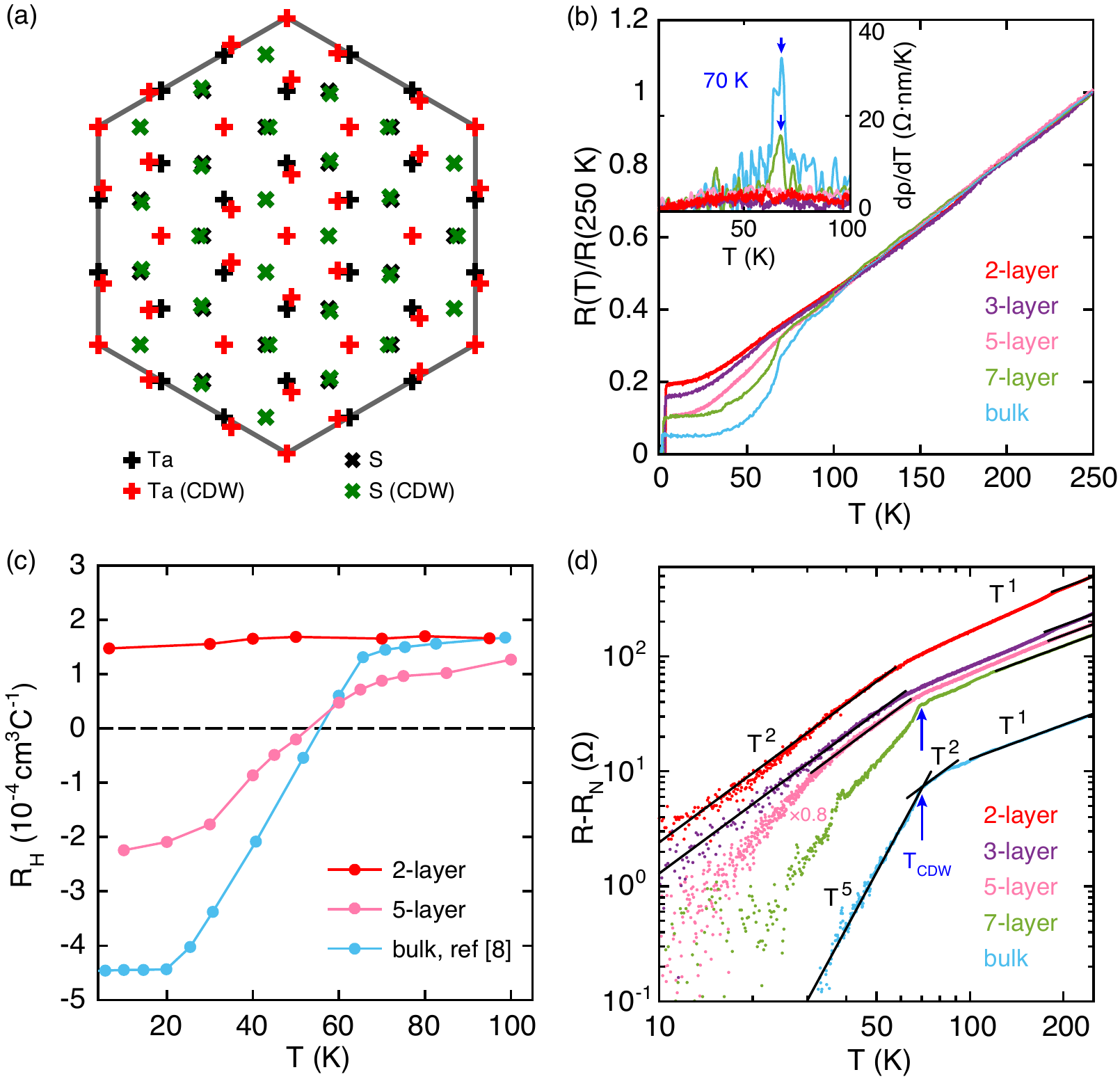}
\caption{
(a) Illustration of the atom position of Ta and S atoms in the normal phase and the CDW phase. The periodicity of the CDW order is $3\times 3$. (b) Normalized resistance $R(T)/R(250 K)$ for 2, 3, 5, 7-layer and bulk samples, measured while cooling down. Inset: derivative of the resistivity $\rho=d\cdot R$ close to the CDW ordering temperature. An arrow is used to mark the T$_{CDW}$ for bulk and 7-layer, which both show a peak in $d\rho/dT$ at 70 K. (c) Hall coefficient $R_H = d\cdot V_H/(I\cdot B)$ measured while cooling down. Data for the bulk crystal is from \cite{thompson1972effects}. (d) Resistance $R-R_N$ as a function of temperature plotted in a log-log scale, where $R_N$ is the residual resistance just above the onset temperature of superconductivity. For clarity, data for 5-layer is scaled by a factor of 0.8.}
\label{fig2}
\end{figure}

In addition to SC transition, bulk 2H-TaS$_2$ is known to manifest CDW order below 70 K.
Fig.~\ref{fig2} (a) illustrates the atomic displacements in the CDW state for monolayer.
Two well established indicators of the CDW phase in bulk TaS$_2$ are a kink in the resistivity occurring at $T_{CDW}$ = 70 K and a change of sign in the Hall coefficient as the temperature is reduced below $T_{CDW}$. 
We find that both disappear as the sample thickness is reduced towards the 2D limit.  
Shown in Fig.~\ref{fig2} (b) is a plot of normalized resistance versus temperature on a linear scale.
All samples manifest a linear decrease of resistance at high temperatures, consistent with phonon limited resistivity in a normal metal \cite{el2013superconductivity}.
Below 70 K, the 7-layer and bulk devices undergo a CDW phase transition, producing a sudden drop in resistance.
This is confirmed by calculating the temperature derivative of the
resistivity, as shown in the inset of Fig.~\ref{fig2} (b).
A peak in $d\rho/dT$ develops close to the transition.
In 2, 3, 5-layer, however, such a sudden drop is not noticeable.

The Hall effect has been used to verify the existence of a phase transition.
The Hall coefficient is found to demonstrate a broad transition between 70 and 20 K and a change in sign at 56 K in bulk crystals \cite{thompson1972effects}.
This indicates that the CDW transition not only induces a structural change, but also alters the electronic properties of the material.
It has been shown that a two-carrier model with light holes and heavy electrons is necessary to explain the opposite signs for the Seebeck and Hall coefficients measured above the CDW transition temperature; a single-carrier model describes the low-temperature behavior \cite{thompson1972effects}.
In Fig.~\ref{fig2} (c), we plot the Hall coefficient of three representative thicknesses below 100 K.
In the 5-layer device, a significant deviation from the bulk behavior is already apparent: the overall magnitude of $R_H$ and temperature where it switches sign, are both diminished. 
However, the most striking fact is the weak temperature dependence and absence of a sign change in the 2-layer sample. 
This provides unambiguous evidence that the CDW transition occurring in bulk samples is absent in the ultra-thin limit.

Another interesting observation is the emergence of $R\sim T^2$ behavior in ultra-thin samples.
In Fig.~\ref{fig2} (d), we plot the subtracted resistance $R-R_N$ on a log-log scale.
For bulk, the linear temperature dependence is disrupted by a sudden switch to $T^5$ near $T_{CDW}$.
$R \sim T^5$ has been well known as a consequence of electron-phonon scattering at temperatures lower than the Debye temperature $\Theta_D$.
In contrast, a gradual transition to $R\sim T^2$ is observed in 2, 3-layer; it persist from 55 K down to the onset temperature of superconductivity.
Naively, such $T^2$ behavior results from electron-electron scattering within Fermi liquid theory. 
However, we find that the observed $T^2$ coefficient is too large to be explained within such a framework. 
Moreover, assuming the e-e scattering strength is not greatly altered from 3D to 2D \cite{SM}, it is implausible that the $T^2$ behavior is completely absent in thick samples in the same temperature range.
Here we propose an alternative mechanism that leads to a $T^2$ resistivity: scattering of electrons by soft phonons, i.e. critical CDW fluctuations, which can happen close to a finite momentum ordering transition \cite{hlubina1995resistivity} (see Supplemental Material \cite{SM}).
This picture is motivated by the longitudinal and Hall resistivity data indicating the disappearance of the CDW order in thin samples. It assumes that although the long range CDW order has been destroyed, strong CDW fluctuations remain, which can scatter electrons.  

It is worth noting that a similar anti-correlation of trends of SC and CDW transition has also been observed in 2H-TaS$_2$ crystals under pressure \cite{freitas2016strong, abdel2016enhancement} and single crystal alloys \cite{Wagner:2008tuning,Fang:2016cs,Li:2017gs}.
To better understand the connection between CDW and superconductivity, we recall that in McMillan's theory, the critical temperature is expressed as \cite{mcmillan1968transition}
\begin{equation} \label{eqMcmillan}
T_c=\frac{\Theta_D}{1.45} \exp[-\frac{1.04(1+\lambda)}{\lambda-\mu^*(1+0.62\lambda)}],
\end{equation}
where $\mu^*$ is the Coulomb pseudopotential of Morel and Anderson, and $\lambda$ is the electron-phonon coupling constant.
Assuming $\mu^*=0.15$ as suggested by McMillan \cite{mcmillan1968transition}, one can evaluate the inverted form of Eq. (\ref{eqMcmillan})
and obtain $\lambda=0.482$ for TaS$_2$ with $\Theta_D=250$~K \cite{Schlicht:2001hn} and $T_c=0.8$~K \cite{NavarroMoratalla:2016df}, indicating that TaS$_2$ lies in the intermediate coupling regime.

\begin{figure}
\includegraphics[width=3.375 in]{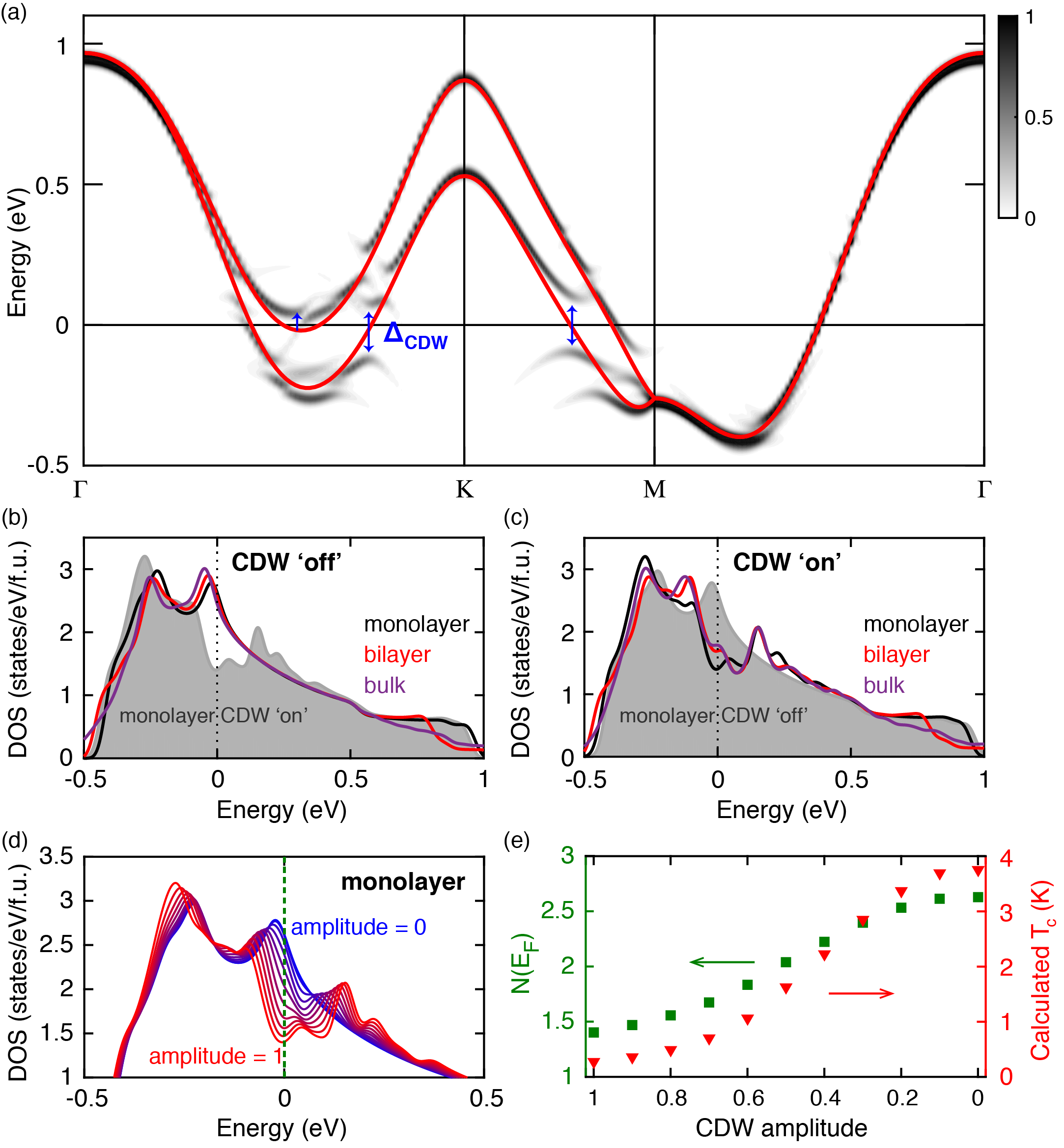}
\caption{(a) Band structure for monolayer 2H-TaS$_2$. The grey lines show the unfolded band structure compared with original band structure in the normal phase (red lines).  (b) Density of states for monolayer/bilayer/bulk in the normal phase. Monolayer in the CDW phase is plotted as a grey shade for reference. (c) Density of states for monolayer/bilayer/bulk in the CDW phase. Monolayer in the normal phase is plotted as a grey shade for reference. (d) A comparison of density of states close to Fermi level for monolayer with various CDW amplitudes ranging from 1 (full amplitude) to 0 (total suppression). (e) Left axis: density of states at Fermi level $N(E_F)$ as a function of CDW amplitude for monolayer. Right axis: $T_c$ from Eq.~(\ref{eqMcmillan}) using calculated $N(E_F)$.
}
\label{fig_theory}
\end{figure}

The CDW instability allows electronic systems to lower their energy by inducing energy gaps in the spectrum.
Since $N(E_F)$ can affect both $\mu^*$ and $\lambda$, it plays an important role in determining $T_c$.
We investigate the electronic and vibrational properties of 2H-TaS$_2$ based on density functional theory (DFT).
First, we implemented in VASP code \cite{kresse1996comput,kresse1996comput2} in order to obtain the DOS in the normal and the CDW phases for monolayer, bilayer and bulk. 
A comparison of Fig.~\ref{fig_theory} (b) and (c) reveals an appreciable reduction of DOS near the Fermi level induced by CDW order for all three thicknesses.
This is consistent with previous magnetic susceptibility and heat capacity experiments showing a sharp drop of density of states below the CDW transition \cite{di1971superconductivity,mattheiss1973band}.
Further, to visualize the effects of CDW on the pristine band structure, we derive the unfolded band structure in the CDW phase, shown in Fig.~\ref{fig_theory} (a), based on the Fourier decompositions of the Bloch wavefunctions from the tight-binding Hamiltonians \cite{farjam2015projection,popescu2012extracting} (see Supplemental Material \cite{SM}).
It is clearly seen that a band gap, $\Delta_{CDW}$, emerges in the inner pocket around \textrm{K} along $\Gamma$-\textrm{K} and \textrm{K}-\textrm{M}.
In addition, the saddle point located along the $\Gamma$-\textrm{K}, is shifted to energies above the Fermi level.
Next, we compute the phonon dispersion for the bulk and monolayer (see Supplemental Material \cite{SM}).
In both the bulk and monolayer, an acoustic mode that involves in-plane motion of Ta atoms softens and becomes unstable as the electronic temperature is lowered. 
We found that in both cases the instability occurs at approximately the same wave vector that corresponds to the CDW ordering $\bm{Q}_{CDW}\approx 2/3~\Gamma$\textrm{M} \cite{albertini2017effect}.

We then investigate the impact of progressive weakening of the CDW with decreasing thickness by varying the magnitude of atomic distortion. 
A scaling factor, from 1 to 0, is used to define the fraction by which the magnitude of the atomic displacement is reduced with respect to the stable distorted configuration.
The corresponding DOS as a function of atom displacement amplitude is calculated and plotted in Fig.~\ref{fig_theory} (d).
Using the bulk as a starting point, we take into account the change in $N(E_F)$ and a small phonon energy shift calculated for monolayer,
and plot the predicted $T_c$ within the McMillan formalism in Fig.~\ref{fig_theory} (e).
An enhancement of $T_c$ up to 3.75 K is achieved when the amplitude of CDW goes to zero.
This gives a rough estimate of the impact of the suppression of the CDW order on the superconducting $T_c$, which leads to a reasonable prediction that aligns with the experimental value.
We note that it is not entirely clear that the enhancement of $T_c$ is solely due to the enhanced DOS. 
There are several factors impacting $T_c$ that have been discussed previously but not addressed in our calculation, such as substrate effects, the presence of a van Hove singularity near the Fermi level and enhanced electron-phonon coupling due to reduced screening in two dimensions \cite{NavarroMoratalla:2016df}, and a weaker interlayer Cooper pairing \cite{xi2016ising}.

This is not the first experiment indicating that a CDW phase transition vanishes in reduced dimensions \cite{thompson1972effects, ayari2007realization, Pan:2017gm, sanders2016crystalline, yoshida2015memristive}.
Its origin is still under debate \cite{ayari2007realization, ge2012effect}.
Recently, a study showed that lattice fluctuations arising from the strong electron-phonon coupling act to suppress the onset temperature of CDW order, leading to a pseudogap phase characterized by local order and strong phase fluctuations \cite{flicker2016charge}.
This is consistent with our model of presence of soft phonons, or CDW fluctuations \cite{naito1982electrical} as primary contributor to the $T^2$ behavior of resistivity observed above $T_{CDW}$.
More interestingly, theory predicts that quantum fluctuations caused by proximity to a CDW transition can boost superconducting pairing by providing sources of bosonic excitations \cite{wang2015enhancement}.
Although there is no direct evidence that CDW fluctuations facilitate superconductivity in 2H-TaS$_2$, this scenario reveals a potentially rich relationship between CDW and SC.

In conclusion, we observe enhanced superconductivity in atomically thin 2H-TaS$_2$ accompanied with suppression of the CDW order. 
Our electronic band structure calculation shows that suppression of the CDW phase leads to a substantial increase in $N(E_F)$, which acts to boost the superconducting $T_c$.
We further suggest that the emergence of $R\sim T^2$ behavior in ultra-thin samples is attributable to the scattering of electrons with soft phonon modes, indicative of critical CDW fluctuations.
Future studies of the layer dependence of the CDW order, for example, STM/STS and ultrafast spectroscopy studies, will be essential to understanding both the origin of the CDW and the relationship between CDW and superconductivity. \\

\begin{acknowledgments}
We thank Yuan Cao, Jason Luo and Jiarui Li for experimental help. 
We also thank Patrick A. Lee, Dennis Huang, Miguel A. Cazalilla and Bertrand I. Halperin for fruitful discussions.
This work has been primarily supported by the US DOE, BES Office, Division of Materials
Sciences and Engineering under Award DE-SC0001819 (YY and PJH) and by the Gordon and Betty Moore FoundationÕs EPiQS Initiative through Grant GBMF4541 to PJH Fabrication work (ENM) and theory analysis were partly supported by the NSF-STC Center for Integrated Quantum Materials under award No. DMR-1231319 (VF, SF) and ARO MURI Award W911NF-14-0247 (EK). 
This work made use of the MRSEC Shared Experimental Facilities supported by NSF under award No. DMR-0819762 and of HarvardÕs CNS, supported by NSF under Grant ECS-0335765. 
SF used Odyssey cluster of the FAS by the Research Computing Group at Harvard University, and the Extreme Science and Engineering Discovery Environment, which is supported by NSF Grant No. ACI-1053575.
JR acknowledges the Gordon and Betty Moore Foundation under the EPiQS initiative under Grant No. GBMF4303. 
A portion of this work was performed at the National High Magnetic Field Laboratory, which is supported by NSF Cooperative Agreement No. DMR-1157490 and the State of Florida.
Growth of hexagonal boron nitride crystals was supported by the Elemental Strategy Initiative conducted by the MEXT, Japan and JSPS KAKENHI Grant Numbers JP15K21722 and JP25106006.
\end{acknowledgments}

\bibliography{Yafang_ref.bib}

\end{document}